\documentclass[10pt,a4paper,onecolumn]{article}
\usepackage{marginnote}
\usepackage{graphicx}
\usepackage[rgb,svgnames]{xcolor}
\usepackage{authblk,etoolbox}
\usepackage{titlesec}
\usepackage{calc}
\usepackage{tikz}
\usepackage[pdfa]{hyperref}
\usepackage{hyperxmp}
\hypersetup{%
    unicode=true,
    pdfapart=3,
    pdfaconformance=B,
    pdftitle={unimpeded: A Public Nested Sampling Database for Bayesian
Cosmology},
    pdfauthor={Dily Duan Yi Ong, Will Handley},
    pdfpublication={Journal of Open Source Software},
    pdfpublisher={Open Journals},
    pdfissn={2475-9066},
    pdfpubtype={journal},
    pdfvolumenum={},
    pdfissuenum={},
    pdfdoi={10.xxxxxx/draft},
    pdfcopyright={Copyright (c) 1970, Dily Duan Yi Ong, Will Handley},
    pdflicenseurl={http://creativecommons.org/licenses/by/4.0/},
    colorlinks=true,
    linkcolor=[rgb]{0.0, 0.5, 1.0},
    citecolor=Blue,
    urlcolor=[rgb]{0.0, 0.5, 1.0},
    breaklinks=true
}
\usepackage{caption}
\usepackage{orcidlink}
\usepackage{tcolorbox}
\usepackage{amssymb,amsmath}
\usepackage{ifxetex,ifluatex}
\usepackage{seqsplit}
\usepackage{xstring}

\usepackage{float}
\let\origfigure\figure
\let\endorigfigure\endfigure
\renewenvironment{figure}[1][2] {
    \expandafter\origfigure\expandafter[H]
} {
    \endorigfigure
}

\usepackage{fixltx2e} 

\NewDocumentCommand\citeproctext{}{}
\NewDocumentCommand\citeproc{mm}{%
  \begingroup\def\citeproctext{#2}\cite{#1}\endgroup}
\makeatletter
 \let\@cite@ofmt\@firstofone
 \def\@biblabel#1{}
 \def\@cite#1#2{{#1\if@tempswa , #2\fi}}
\makeatother
\newlength{\cslhangindent}
\newlength{\csllabelwidth}
\newenvironment{CSLReferences}[2] 
 {\begin{list}{}{%
  \setlength{\itemindent}{0pt}
  \setlength{\leftmargin}{0pt}
  \setlength{\parsep}{0pt}
  \ifodd #1
   \setlength{\leftmargin}{\cslhangindent}
   \setlength{\itemindent}{-1\cslhangindent}
  \fi
  \setlength{\itemsep}{#2\baselineskip}}}
 {\end{list}}
\usepackage{calc}

\usepackage[top=3.5cm, bottom=3cm, right=1.5cm, left=1.0cm,
            headheight=2.2cm, reversemp, includemp, marginparwidth=4.5cm]{geometry}

\titleformat{\section}
  {\normalfont\sffamily\Large\bfseries}
  {}{0pt}{}
\titleformat{\subsection}
  {\normalfont\sffamily\large\bfseries}
  {}{0pt}{}
\titleformat{\subsubsection}
  {\normalfont\sffamily\bfseries}
  {}{0pt}{}
\titleformat*{\paragraph}
  {\sffamily\normalsize}

\usepackage{fancyhdr}
\makeatletter
\let\ps@plain\ps@fancy
\fancyheadoffset[L]{4.5cm}
\fancyfootoffset[L]{4.5cm}

\definecolor{linky}{rgb}{0.0, 0.5, 1.0}

\newtcolorbox{repobox}
   {colback=red, colframe=red!75!black,
     boxrule=0.5pt, arc=2pt, left=6pt, right=6pt, top=3pt, bottom=3pt}

\newcommand{\ExternalLink}{%
   \tikz[x=1.2ex, y=1.2ex, baseline=-0.05ex]{%
       \begin{scope}[x=1ex, y=1ex]
           \clip (-0.1,-0.1)
               --++ (-0, 1.2)
               --++ (0.6, 0)
               --++ (0, -0.6)
               --++ (0.6, 0)
               --++ (0, -1);
           \path[draw,
               line width = 0.5,
               rounded corners=0.5]
               (0,0) rectangle (1,1);
       \end{scope}
       \path[draw, line width = 0.5] (0.5, 0.5)
           -- (1, 1);
       \path[draw, line width = 0.5] (0.6, 1)
           -- (1, 1) -- (1, 0.6);
       }
   }

\definecolor{c53baa1}{RGB}{83,186,161}
\definecolor{c202826}{RGB}{32,40,38}

\patchcmd{\@maketitle}{center}{flushleft}{}{}
\patchcmd{\@maketitle}{center}{flushleft}{}{}
\patchcmd{\@maketitle}{\LARGE}{\LARGE\sffamily}{}{}
\def\maketitle{{%
  
  \AB@maketitle}}
\renewcommand\AB@affilsepx{ \protect\Affilfont}
\renewcommand\AB@affilnote[1]{{\bfseries #1}\hspace{3pt}}
\renewcommand{\affil}[2][]%
   {\newaffiltrue\let\AB@blk@and\AB@pand
      \if\relax#1\relax\def\AB@note{\AB@thenote}\else\def\AB@note{#1}%
        \setcounter{Maxaffil}{0}\fi
        \begingroup
        \let\href=\href@Orig
        \let\protect\@unexpandable@protect
        \def\thanks{\protect\thanks}\def\footnote{\protect\footnote}%
        \@temptokena=\expandafter{\AB@authors}%
        {\def\\{\protect\\\protect\Affilfont}\xdef\AB@temp{#2}}%
         \xdef\AB@authors{\the\@temptokena\AB@las\AB@au@str
         \protect\\[\affilsep]\protect\Affilfont\AB@temp}%
         \gdef\AB@las{}\gdef\AB@au@str{}%
        {\def\\{, \ignorespaces}\xdef\AB@temp{#2}}%
        \@temptokena=\expandafter{\AB@affillist}%
        \xdef\AB@affillist{\the\@temptokena \AB@affilsep
          \AB@affilnote{\AB@note}\protect\Affilfont\AB@temp}%
      \endgroup
       \let\AB@affilsep\AB@affilsepx
}
\makeatother

\renewcommand\Affilfont{\sffamily\small\mdseries}
\ifnum 0\ifxetex 1\fi\ifluatex 1\fi=0 
  \usepackage[T1]{fontenc}
  \usepackage[utf8]{inputenc}

\else 
  \ifxetex
    \usepackage{mathspec}
    \usepackage{fontspec}

  \else
    \usepackage{fontspec}
  \fi
  \defaultfontfeatures{Scale=MatchLowercase}
  \defaultfontfeatures[\sffamily]{Ligatures=TeX}
  \defaultfontfeatures[\rmfamily]{Ligatures=TeX,Scale=1}

\fi
\IfFileExists{upquote.sty}{\usepackage{upquote}}{}
\IfFileExists{microtype.sty}{%
\usepackage{microtype}
\UseMicrotypeSet[protrusion]{basicmath} 
}{}

\PassOptionsToPackage{usenames,dvipsnames}{color} 
\ifLuaTeX
\usepackage[bidi=basic]{babel}
\else
\usepackage[bidi=default]{babel}
\fi
\babelprovide[main,import]{american}

\def\languageshorthands#1{}
\usepackage{color}
\usepackage{fancyvrb}

\DefineVerbatimEnvironment{Highlighting}{Verbatim}{commandchars=\\\{\}}
\newenvironment{Shaded}{}{}

\newcommand{\CommentTok}[1]{\textcolor[rgb]{0.38,0.63,0.69}{\textit{#1}}}

\newcommand{\DecValTok}[1]{\textcolor[rgb]{0.25,0.63,0.44}{#1}}

\newcommand{\ImportTok}[1]{\textcolor[rgb]{0.00,0.50,0.00}{\textbf{#1}}}

\newcommand{\NormalTok}[1]{#1}
\newcommand{\OperatorTok}[1]{\textcolor[rgb]{0.40,0.40,0.40}{#1}}

\newcommand{\StringTok}[1]{\textcolor[rgb]{0.25,0.44,0.63}{#1}}

\usepackage{graphicx}
\makeatletter
\newsavebox\pandoc@box
\newcommand*\pandocbounded[1]{
  \sbox\pandoc@box{#1}%
  \Gscale@div\@tempa{\textheight}{\dimexpr\ht\pandoc@box+\dp\pandoc@box\relax}%
  \Gscale@div\@tempb{\linewidth}{\wd\pandoc@box}%
  \ifdim\@tempb\p@<\@tempa\p@\let\@tempa\@tempb\fi
  \ifdim\@tempa\p@<\p@\scalebox{\@tempa}{\usebox\pandoc@box}%
  \else\usebox{\pandoc@box}%
  \fi%
}
\def\fps@figure{htbp}
\makeatother

\usepackage{graphicx,grffile}
\makeatletter
\def\maxwidth{\ifdim\Gin@nat@width>\linewidth\linewidth\else\Gin@nat@width\fi}
\def\maxheight{\ifdim\Gin@nat@height>\textheight\textheight\else\Gin@nat@height\fi}
\makeatother
\setkeys{Gin}{width=\maxwidth,height=\maxheight,keepaspectratio}
\IfFileExists{parskip.sty}{%
\usepackage{parskip}
}{
\setlength{\parindent}{0pt}
\setlength{\parskip}{6pt plus 2pt minus 1pt}
}
\providecommand{\tightlist}{%
  \setlength{\itemsep}{0pt}\setlength{\parskip}{0pt}}
\ifx\paragraph\undefined\else
\let\oldparagraph\paragraph
\renewcommand{\paragraph}[1]{\oldparagraph{#1}\mbox{}}
\fi
\ifx\subparagraph\undefined\else
\let\oldsubparagraph\subparagraph
\renewcommand{\subparagraph}[1]{\oldsubparagraph{#1}\mbox{}}
\fi
\ifLuaTeX
  \usepackage{selnolig}  
\fi

\title{unimpeded: A Public Nested Sampling Database for Bayesian
Cosmology}

\author[1,2,3%
\ensuremath\mathparagraph]{Dily Duan Yi Ong%
  \,\orcidlink{0009-0004-8688-5088}\,%
}
\author[1,2%
]{Will Handley%
  \,\orcidlink{0000-0002-5866-0445}\,%
}

\affil[1]{Kavli Institute for Cosmology, Madingley Road, Cambridge, CB3
0HA, UK%
}
\affil[2]{Astrophysics Group, Cavendish Laboratory, J.J. Thomson Avenue,
Cambridge, CB3 0HE, UK%
}
\affil[3]{Newnham College, Sidgwick Avenue, Cambridge, CB3 9DF, UK%
}
\affil[$\mathparagraph$]{Corresponding author}
\date{\vspace{-2.5ex}}

\begin{document}
\maketitle

\marginpar{

  \begin{flushleft}
  \sffamily\small

  {\bfseries DOI:} \href{https://doi.org/10.xxxxxx/draft}{\color{linky}{10.xxxxxx/draft}}

  \vspace{2mm}
    {\bfseries Software}
  \begin{itemize}
    \setlength\itemsep{0em}
    \item \href{https://github.com/openjournals}{\color{linky}{Review}} \ExternalLink
    \item \href{https://github.com/openjournals}{\color{linky}{Repository}} \ExternalLink
    \item \href{https://doi.org/10.5281}{\color{linky}{Archive}} \ExternalLink
  \end{itemize}

  \vspace{2mm}
  
    \par\noindent\hrulefill\par

  \vspace{2mm}

  {\bfseries Editor:} \href{https://joss.theoj.org}{Open
Journals} \ExternalLink
   \\
  \vspace{1mm}
    {\bfseries Reviewers:}
  \begin{itemize}
  \setlength\itemsep{0em}
    \item \href{https://github.com/openjournals}{@openjournals}
    \end{itemize}
    \vspace{2mm}
  
    {\bfseries Submitted:} unsubmitted\\
    {\bfseries Published:} unpublished

  \vspace{2mm}
  {\bfseries License}\\
  Authors of papers retain copyright and release the work under a Creative Commons Attribution 4.0 International License (\href{https://creativecommons.org/licenses/by/4.0/}{\color{linky}{CC BY 4.0}}).

  \end{flushleft}
}

\section{Summary}\label{summary}

Bayesian inference is central to modern cosmology. While parameter
estimation is achievable with unnormalised posteriors traditionally
obtained via MCMC methods, comprehensive model comparison and tension
quantification require Bayesian evidences and normalised posteriors,
which remain computationally prohibitive for many researchers. To
address this, we present \texttt{unimpeded}, a publicly available Python
library and data repository providing DiRAC-funded (DP192 and 264)
pre-computed nested sampling and MCMC chains with their normalised
posterior samples, computed using \texttt{Cobaya}
(\citeproc{ref-Torrado2021}{Torrado \& Lewis, 2021}) and the Boltzmann
solver CAMB (\citeproc{ref-Lewis1999}{Lewis et al., 2000};
\citeproc{ref-Lewis2002}{Lewis \& Bridle, 2002}). \texttt{unimpeded}
delivers systematic analysis across a grid of eight cosmological models
(including ΛCDM and seven extensions) and 39 modern cosmological
datasets (comprising individual probes and their pairwise combinations).
The built-in tension statistics calculator enables rapid computation of
six tension quantification metrics. All chains are hosted on
Zenodo\footnote{https://zenodo.org/} with permanent access via the
\texttt{unimpeded} API, analogous to the renowned Planck Legacy Archive
(\citeproc{ref-Dupac2015}{Dupac et al., 2015}) but utilising nested
sampling (\citeproc{ref-Skilling2006}{Skilling, 2006}) in addition to
traditional MCMC methods.

\section{\texorpdfstring{\texttt{unimpeded}}{unimpeded}}\label{unimpeded}

\texttt{unimpeded} addresses these challenges directly. It provides a
pip-installable tool that leverages the \texttt{anesthetic} package
(\citeproc{ref-Handley2019}{W. Handley, 2019}) for analysis and
introduces a seamless Zenodo integration for data management. The nested
sampling theory and methodology are detailed in
(\citeproc{ref-Ong2025}{Ong \& Handley, 2025}). Its main features are:

\begin{enumerate}
\def\labelenumi{\arabic{enumi}.}
\tightlist
\item
  \textbf{A Public Nested Sampling Grid:} The package provides access to
  a pre-computed grid of nested sampling chains and MCMC chains for 8
  cosmological models (standard \(\Lambda\)CDM and seven extensions),
  run against 39 datasets (comprising individual probes and their
  pairwise combinations). This saves the community significant
  computational resources and provides a common baseline for new
  analyses. Evidence and Kullback-Leibler divergence can be calculated
  jointly with \texttt{anesthetic} for model comparison and quantifying
  the constraining power of datasets and models, respectively. The
  scientific results from this grid are presented in
  (\citeproc{ref-Ong2025}{Ong \& Handley, 2025}).
\item
  \textbf{Archival and Reproducibility via Zenodo:} \texttt{unimpeded}
  automates the process of archiving analysis products. The
  \texttt{DatabaseCreator} class bundles chains and metadata, uploading
  them to a Zenodo community to generate a permanent, citable Digital
  Object Identifier (DOI). The \texttt{DatabaseExplorer} class allows
  public user to easily download and analyse these chains, promoting
  open science and effortless reproducibility. Figure 1 illustrates the
  \texttt{unimpeded} ecosystem, detailing its three core functions. For
  data generation, it configures YAML files for HPC nested sampling. It
  then archives the chains on Zenodo, ensuring reproducibility with
  permanent DOIs, and finally provides an interface for post-processing
  analysis and visualisation with \texttt{anesthetic}. The following
  example demonstrates downloading chains:
\end{enumerate}

\begin{Shaded}
\begin{Highlighting}[]
\ImportTok{from}\NormalTok{ unimpeded.database }\ImportTok{import}\NormalTok{ DatabaseExplorer}

\CommentTok{\# Initialise DatabaseExplorer}
\NormalTok{dbe }\OperatorTok{=}\NormalTok{ DatabaseExplorer()}

\CommentTok{\# Get a list of currently available models and datasets}
\NormalTok{models\_list }\OperatorTok{=}\NormalTok{ dbe.models}
\NormalTok{datasets\_list }\OperatorTok{=}\NormalTok{ dbe.datasets}

\CommentTok{\# Choose model, dataset and sampling method}
\NormalTok{method }\OperatorTok{=} \StringTok{\textquotesingle{}ns\textquotesingle{}}  \CommentTok{\# \textquotesingle{}ns\textquotesingle{} for nested sampling, \textquotesingle{}mcmc\textquotesingle{} for MCMC}
\NormalTok{model }\OperatorTok{=} \StringTok{"klcdm"}  \CommentTok{\# from models\_list}
\NormalTok{dataset }\OperatorTok{=} \StringTok{"des\_y1.joint+planck\_2018\_CamSpec"}  \CommentTok{\# from datasets\_list}

\CommentTok{\# Download samples chain}
\NormalTok{samples }\OperatorTok{=}\NormalTok{ dbe.download\_samples(method, model, dataset)}

\CommentTok{\# Download Cobaya and PolyChord run settings}
\NormalTok{info }\OperatorTok{=}\NormalTok{ dbe.download\_info(method, model, dataset)}
\end{Highlighting}
\end{Shaded}

\begin{enumerate}
\def\labelenumi{\arabic{enumi}.}
\tightlist
\item
  \textbf{Tension Statistics Calculator:} With the nested sampling
  chains and the built-in tension statistics calculator, six tension
  quantification metrics with different characteristics are available,
  including the \(R\) statistic, information ratio \(I\), suspiciousness
  \(S\), Gaussian model dimensionality \(d_G\), tension significance in
  units of \(\sigma\), and p-value. Each of them has unique
  characteristics optimised for different tasks, thoroughly discussed in
  (\citeproc{ref-Ong2025}{Ong \& Handley, 2025}). \texttt{unimpeded}
  implements these statistics with the necessary correction to account
  for discarded prior volume (\citeproc{ref-Handley2019a}{W. Handley \&
  Lemos, 2019}; \citeproc{ref-Handley2021}{Will Handley \& Lemos, 2021};
  \citeproc{ref-Ong2025}{Ong \& Handley, 2025}). Figure 2 demonstrates
  the tension calculator output showing p-value derived tension
  significance (\(\sigma\)) for 31 pairwise dataset combinations across
  8 cosmological models, sorted by significance to highlight the
  datasets in tension. Caution should be exercised when combining them.
  The following minimal example demonstrates the usage:
\end{enumerate}

\begin{Shaded}
\begin{Highlighting}[]
\ImportTok{from}\NormalTok{ unimpeded.tension }\ImportTok{import}\NormalTok{ tension\_calculator}

\NormalTok{tension\_samples }\OperatorTok{=}\NormalTok{ tension\_calculator(method}\OperatorTok{=}\StringTok{\textquotesingle{}ns\textquotesingle{}}\NormalTok{,}
\NormalTok{                                      model}\OperatorTok{=}\StringTok{\textquotesingle{}lcdm\textquotesingle{}}\NormalTok{,}
\NormalTok{                                      datasetA}\OperatorTok{=}\StringTok{\textquotesingle{}planck\_2018\_CamSpec\textquotesingle{}}\NormalTok{,}
\NormalTok{                                      datasetB}\OperatorTok{=}\StringTok{\textquotesingle{}des\_y1.joint\textquotesingle{}}\NormalTok{,}
\NormalTok{                                      nsamples}\OperatorTok{=}\DecValTok{1000}\NormalTok{)}
\end{Highlighting}
\end{Shaded}

\begin{figure}
\centering
\pandocbounded{\includegraphics[keepaspectratio]{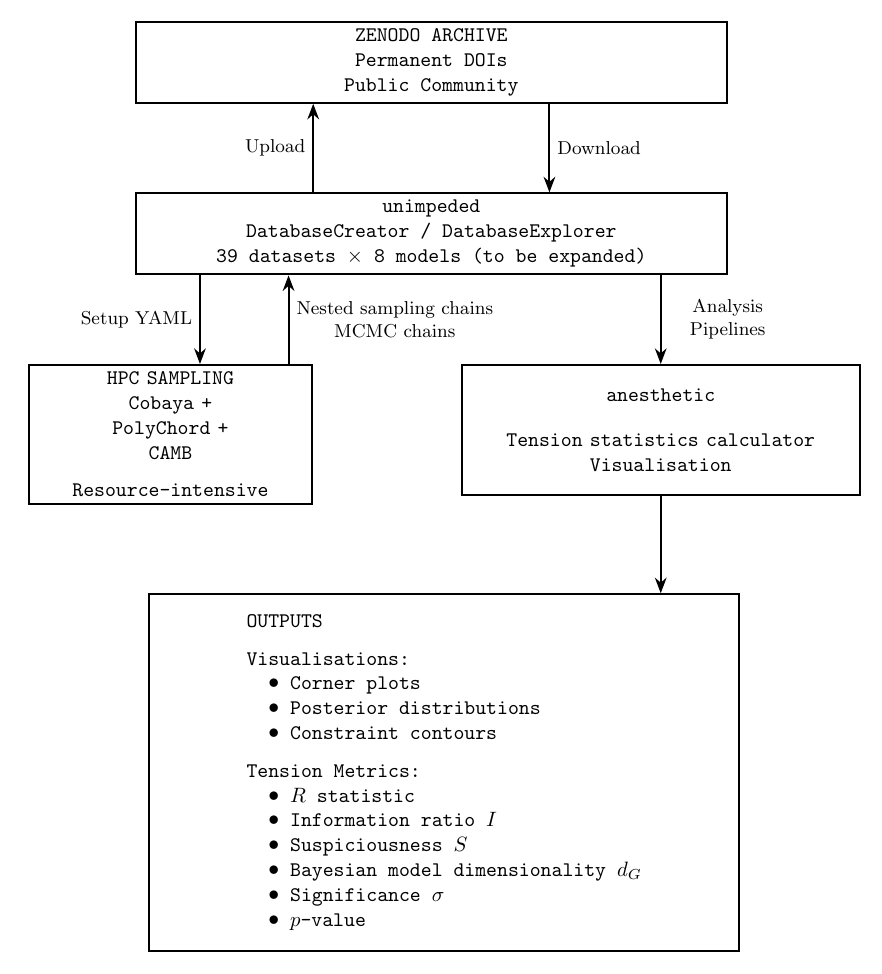}}
\caption{The \texttt{unimpeded} ecosystem and workflow. At the centre,
\texttt{unimpeded} manages data archival and retrieval through Zenodo,
providing permanent DOIs and public access to pre-computed chains. For
data generation, \texttt{unimpeded} configures YAML files for
resource-intensive HPC nested sampling using \texttt{Cobaya},
\texttt{PolyChord}, and \texttt{CAMB}. For analysis, users download
chains via \texttt{DatabaseExplorer} and leverage \texttt{anesthetic}
for visualisation (corner plots, posterior distributions, constraint
contours) and tension quantification (six metrics: \(R\) statistic,
information ratio \(I\), suspiciousness \(S\), Bayesian model
dimensionality \(d_G\), significance \(\sigma\), and
\(p\)-value).\label{fig:workflow}}
\end{figure}

\begin{figure}
\centering
\pandocbounded{\includegraphics[keepaspectratio]{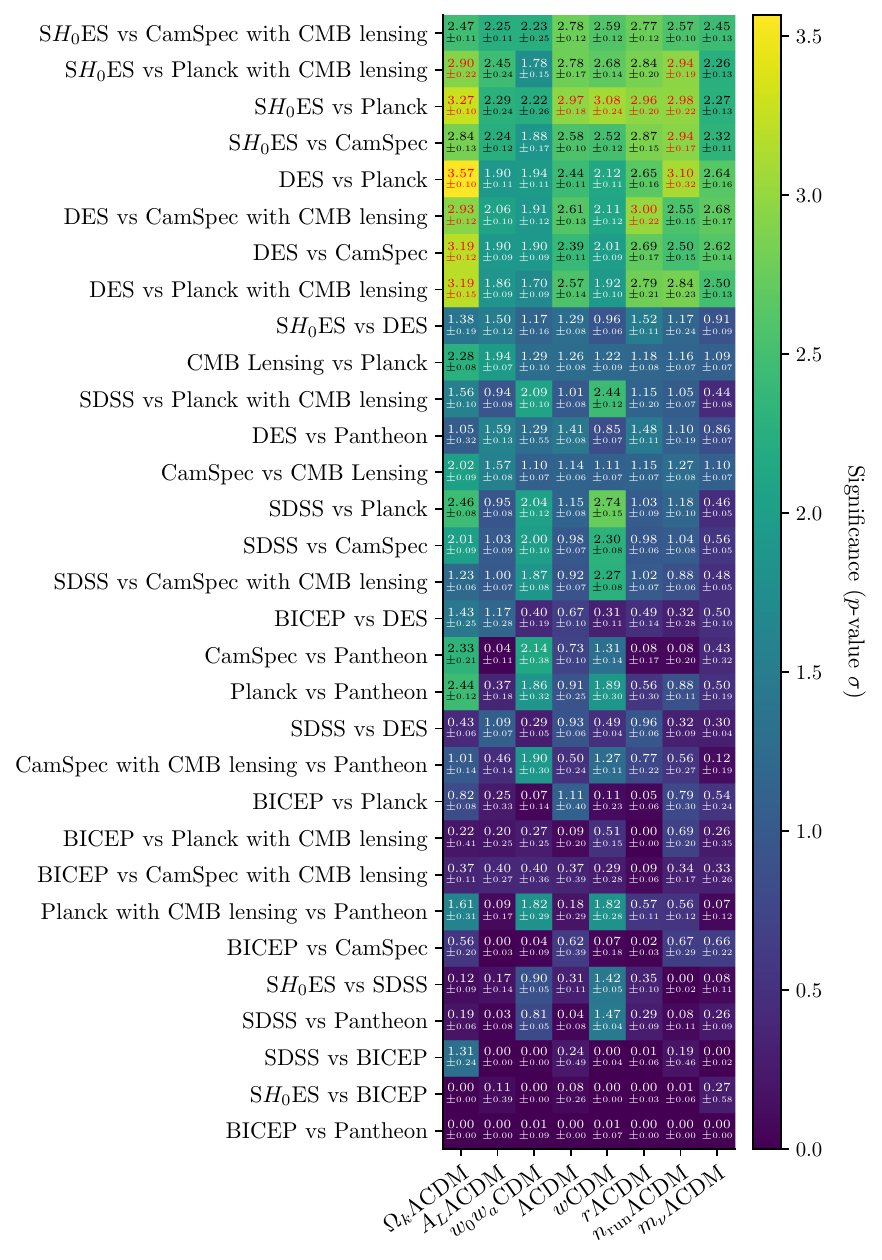}}
\caption{Tension analysis heatmap produced by \texttt{unimpeded} and
\texttt{anesthetic} displaying p-value derived tension significance
(\(\sigma\) values) for 31 pairwise dataset combinations across 8
cosmological models. Rows are sorted by significance, with the most
problematic dataset pairs (highest tension) at the top. This
demonstrates \texttt{unimpeded}'s capability to systematically quantify
tensions and their model dependence.\label{fig:tension_heatmap}}
\end{figure}

While tools like \texttt{getdist} (\citeproc{ref-Lewis2019}{Lewis,
2019}) are excellent for MCMC analysis, and frameworks like
\texttt{CosmoSIS} (\citeproc{ref-Zuntz2015}{Zuntz et al., 2015}) or
\texttt{MontePython} (\citeproc{ref-Brinckmann2019}{Brinckmann \&
Lesgourgues, 2019}) are used for running simulations with samplers like
\texttt{Cobaya} (\citeproc{ref-Torrado2021}{Torrado \& Lewis, 2021}),
\texttt{unimpeded} fills a unique niche. It is not a sampler but a
high-level analysis and database management tool that extends the
capabilities of its underlying engine, \texttt{anesthetic}, to create a
public, reproducible, and statistically robust nested sampling resource
for the cosmology community.

The package is fully documented, tested, and available for installation
via the Python Package Index (PyPI). A Jupyter notebook tutorial is also
available to help users get started.

\section{Acknowledgements}\label{acknowledgements}

We thank the developers of the open-source packages that this work
relies upon, including \texttt{anesthetic}, \texttt{numpy},
\texttt{scipy}, \texttt{pandas}, and \texttt{corner.py}. This work was
performed using the Cambridge Service for Data Driven Discovery (CSD3),
operated by the University of Cambridge Research Computing Service,
provided by Dell EMC and Intel using Tier-2 funding from the Engineering
and Physical Sciences Research Council (capital grant EP/P020259/1), and
DiRAC funding from the Science and Technology Facilities Council
(www.dirac.ac.uk).

\section*{References}\label{references}
\addcontentsline{toc}{section}{References}

\phantomsection\label{refs}

\end{document}